# THE EVOLUTION OF MASSIVE AND VERY MASSIVE STARS IN CLUSTERS

**Dany Vanbeveren**

*Astrophysical Institute, Vrije Universiteit Brussel, 1050 Brussels, Belgium;
dvbevere@vub.ac.be*

*and*

*Leuven Engineering College, Association K.U. Leuven, 3000 Leuven, Belgium;
dany.vanbeveren@groept.be*

**Abstract**

The present paper reviews massive star (initial mass $\leq 120$ M$_\odot$) and very massive star (initial mass $> 120$ M$_\odot$) evolution. I will focus on evolutionary facts and questions that may critically affect predictions of population and spectral synthesis of starburst regions. We discuss the ever-lasting factor 2 or more uncertainty in the stellar wind mass loss rates. We may ask ourselves if stellar rotation is one of the keys to understand the universe, why so many massive stars are binary components and why binaries are ignored or are considered as the poor cousins by some people? And finally, do ultra luminous X-ray sources harbor an intermediate mass black hole with a mass of the order of 1000 M$_\odot$?

## 1. Introduction

The sixties /early seventies are characterized by an explosion of studies of massive star evolution but mainly in the framework of interacting binaries. Two main reasons: the rise of X-ray binary astronomy and the interpretation of Wolf-Rayet (WR) stars as stars that lost their hydrogen rich envelope. In this period the only process one could think of capable to remove a large amount of mass from a star was the Roche lobe overflow (RLOF) in binaries and it was therefore believed that all WR stars were binary members (Kuhi, 1973). However, the Copernicus survey by Snow and Morton (1976) showed that all stars brighter than $M_{bol} = -6$ are losing mass by stellar wind. Subsequent observations revealed that the wind rates were large enough to affect the overall evolution of a massive star, may be large enough in order to explain the existence of single WR stars (the '*Conti scenario*' for single WR stars, Conti, 1976) and the massive single star evolution became hot topic. Furthermore it was demonstrated by Vanbeveren and Conti (1980) that at most 40% of the WR stars are real binary members (a percentage that still holds today), and this of course added to the importance of the Conti scenario.

Since 1990 large massive star evolutionary data sets have been calculated by many groups and it became reasonable in order to simulate the evolution of massive star populations in regions where star formation is a continuous process and in starbursts: population number synthesis and population spectral synthesis became very popular. Models have been published of clusters with single stars only (e.g. Arnault et al., 1989; Maeder, 1991; Mas-Hesse & Kunth 1991; Cervino & Mas-Hesse, 1994; Meynet, 1995; Leitherer & Heckman, 1995; Leitherer et al. 1999) and, more realistically, with a mixture of single stars and close binaries (Yungelson & Tutukov, 1991; Dalton & Sarazin, 1995; Cervino et al., 1996; Schaerer & Vacca, 1998; Pols et al 1991; Van Bever & Vanbeveren, 1997, 2000, 2003; Vanbeveren et al., 1997, 1998a, 1998b, 1998c, 2007; Belczynski et al., 2002).

To understand the chemical evolution of galaxies, massive stars play a very important role. Note however that most existing chemical evolution codes only account for the effects of massive single stars and massive binaries are ignored. Of course, binaries make life (and computer codes) very complex. To my knowledge the only galactic code with a reasonable mixture of single stars and binaries is the Brussels code (for a review and a full description of the code ingredients, see De Donder & Vanbeveren, 2004).

Massive star evolution is subject to uncertainties in the physics of processes that determine their evolution. We distinguish those in the field of stellar wind mass loss in single stars and binary components, in the Roche lobe overflow process in binaries and the related mass and angular momentum transfer between two binary components, in the effect of rotation on stellar evolution. In the first part of the present paper I will critically discuss these uncertainties, and I will try to evaluate the consequences for population synthesis.

Due to the process of mass segregation, massive stars that are born in a dense cluster tend to sink into the cluster core where they may physically collide forming a stellar merger (e.g., Portegies Zwart et al., 1999). The structure of such a merger may be quite different from that of a canonical star. Even more, a merger which may obviously be more massive than the other massive stars, may initiate a chain reaction where more massive stars interact gravitationally with the merger and eventually a runaway merger (= a collision of many massive stars on a short timescale, short compared to the evolutionary timescale of a massive star which is of the order of a few Myr) forms. In the second part of the present paper, we discuss the formation and evolution of these runaway mergers and focus on the uncertainties there as well.

## 2. The evolution of massive stars: facts and uncertainties

Let us start with some general facts.

- The more and the better the observations, the more massive stars are found to be binary components. Accounting for all possible biases, the statement 'most of the massive stars are born as binary components' becomes more and more probable. It is inherent to the overall evolutionary scenario of massive binaries that many merge and form single stars or that many binary components become single after the supernova explosion of its companion. This means that a region where all massive stars are born as binary components will become populated by massive single stars due to binary evolution. Note thus that massive single stars that are observed in a population may have had a binary past implying that their evolution may be very different from the evolution of a canonical single star.
- The observed massive binaries are very often short period binaries with periods of the order of a few days to a few weeks. I have the impression that the period distribution of massive binaries is distinctly different from the period distribution of lower (low) mass binaries and that the average period of massive binaries is smaller than the average period of the lower mass binaries.
- Vanbeveren (1982) proposed a relation between the total mass of the cluster and the maximum mass $M_{max}$ of the stars in that cluster. If the relation is accepted then it was shown in the same paper that the initial mass function (IMF) of clusters should be top heavy compared to the canonical IMF of galactic regions containing many clusters with a typical cluster total mass function. This 1982-proposal has been discussed over again by Weidner and Kroupa (2005) who reached similar conclusions. It is obvious to understand that if indeed there is a relation between the total cluster mass and the maximum mass of the cloud fragment that may become a star, the hypothesis that depending on the angular momentum of the cloud fragment it may either become a single star or a binary implies that there may also be a relation between the total cluster mass and the overall cluster population of binaries with primary mass larger than $M_{max}/2$.

## 2.1. Stellar wind mass loss rates and massive star evolution

The present state of the physics and observations of hot massive star stellar winds has recently been reviewed by Puls et al. (2008) and the results of the studies on the effects of winds on massive star evolution can be summarized as follows.

- The mass loss rates of O-type stars with an initial mass larger than 30-40 $M_\odot$ are uncertain by at least a factor of two (especially the effects of wind inhomogeneities = clumping is poorly known), an uncertainty that exists already a few decades, and unfortunately this factor decides whether or not these rates are important for the evolution in this mass range.
- When the area of the Luminous Blue Variables (LBV) in the HR-diagram is compared to theoretical evolution, we conclude that they are end-core-hydrogen-burning (CHB) or hydrogen-shell-burning stars originating from stars with an initial mass larger than 30-40 $M_\odot$. LBVs have strong stellar winds and they temporarily eject large amounts of mass on a short timescale (e.g., the 19$^{th}$ century event of η Car). The stellar wind rates and the reason for the ejections are very poorly known and this is also the case for their effect on the evolution of stars in the >30-40 $M_\odot$ mass range.
- Stellar evolution in this mass range becomes much less uncertain if one translates the observational fact that there are no red supergiants (RSG) with an initial mass > 30-40 $M_\odot$ into the following condition: 'during the O-type phase and the LBV phase a star with an initial mass > 30-40 $M_\odot$ has to lose enough mass in order to prohibit redward evolution.' Obviously, when such a star is a component of a Case B or C binary, RLOF will not happen (the 'LBV scenario' of massive binaries as introduced by Vanbeveren, 1991).

- Stellar wind mass loss during CHB and hydrogen shell burning is relatively unimportant for the overall evolution of massive stars with initial mass < 30-40 Mo. Stars in this mass range become RSGs and therefore RSG stellar wind dominates their further evolution. Most of the stellar evolutionary codes use the RSG wind formalism proposed by de Jager et al. (1988). However, Vanbeveren et al. (2007) showed that in the luminosity interval $4 < \mathrm{Log}\, L/L_\odot < 5.5$ the latter formalism may significantly underestimate the true rates. This has important consequences for the theoretically predicted population of WR stars in galaxies and in starburst regions. Notice that the effects of this RSG stellar wind may be very important in binaries with a period that is large enough so that this RSG stellar wind operates before the RLOF (and common envelope) phase starts (the RSG scenario of massive binaries as introduced in Vanbeveren, 1996).
- Massive stars that have lost their hydrogen rich layers become WR stars and their further evolution is dominated by WR-like stellar winds. I think that it is still optimistic to state that we know the WR winds within a factor of 2. Unfortunately, once more this factor 2 critically decides upon the theoretically predicted population of WR stars.

## 2.2. Rotation and massive single star evolution

The pre-supernova evolution of rotating massive stars has been investigated intensively the last decade, mainly by the teams in Geneva and Utrecht. Note that including rotation and rotational mixing in stellar structure equations is not easy. Therefore, it is understandable that a team who found the courage to do the hard labor hopes that the effect is enormous. Let us critically discuss if this is the case.

The effects of rotation on stellar evolution can be summarized in three points:

- Rotating stars make larger convective cores compared to non-rotating stars, e.g. convective overshooting (a process of the old days) mimics the effect of rotation on convective cores.
- Rotation induces mixing and transport in regions outside the convective core by meridional circulation and shear instabilities. It is capable to change the chemical composition of these regions in general, of the stellar surface layers in particular, and therefore surface chemical effects predicted by rotation can be compared to observations. Note however that the effect of this mixing on the overall evolution of a massive star is rather small.
- Rotation may enhance the stellar wind mass loss rate. The formalism proposed by Maeder and Meynet (2000) implies that the influence of rotation is marginal for most of the stars with not too large Eddington factor $\Gamma$ (say stars with an initial mass < 40 $M_\odot$. For larger $\Gamma$ factors (say for stars with an initial mass > 40 $M_\odot$), the formalism diverges and this is unrealistic of course (Puls et al., 2008). Since there are no full 2-D hydro-dynamical calculations for rotating stars in this mass range, it is not known in how far the formalism predicts the correct rates. Moreover, the formalism relates the mass loss rate of a rotating star to the mass loss rate of a non-rotating star, and therefore, any uncertainty in the latter implies the same uncertainty in the former.

Meynet and Maeder (2003) investigated the effect of rotation on galactic massive star evolution. The authors compare the evolution of non-rotating stars with those of rotating stars. The initial rotation velocity $v_{rot}$ on the ZAMS is assumed to be 300 km/s because the authors argue that this value predicts an average MS $v_{rot}$ of massive stars of 200-240 km/s as observed (we will discuss this argument later). The results can be summarized as follows:

- major differences occur in the M > 40 $M_\odot$ range, and this is not in the least due to the adopted formalism that describes the effect of rotation on the mass loss rates. Since the formalism is most uncertain in this mass range, hydro-dynamical simulations are urgently needed to confirm the validity of the formalism,
- in the M ≤ 40 $M_\odot$ area, the effect of rotation is much less than in the larger mass range. Stars evolve at slightly larger luminosity ($\Delta \log L/L_\odot \leq 0.1$) mainly because rotating stars have larger convective cores, however it is doubtful whether this will affect the overall spectral synthesis of a massive starburst once the stars with M > 40 $M_\odot$ disappeared,
- It is known for quite some time now that in order to bring the theoretically predicted single WR star population of the Solar neighborhood in better agreement with the observed one, it is necessary that also stars with an initial mass between 25 and 40 $M_\odot$ lose enough mass to become WR stars. Since rotation may enhance the stellar wind mass loss rate, rotation was proposed by Maeder and Meynet (2003) in order to improve the agreement. However, accounting for the uncertain RSG mass loss rates (see before) rotation may not be the only way out (Vanbeveren et. al., 2007).

Let us have a closer look at the observed rotational velocities of massive stars and check the statement that massive stars should initially have an average rotational velocity of 300 km/s. Using the data of Penny (1996) and the method outlined by Lucy (1974) in order to transform a $v_{rot} sini$ distribution into a $v_{rot}$ distribution, Figure 1 gives the $v_{rot}$ distribution of galactic O-type stars (Vanbeveren et al., 1998). We see that a majority of the O-type stars are relatively slow rotators (average $v_{rot}$ = 100-125 km/s). When this is translated into an initial average $v_{rot}$ on the ZAMS of O-type stars, we conclude that this value is < 200 km/s rather than the 300 km/s proposed by Maeder and Meynet (2002). I leave it to the interested reader to check the literature and reach the conclusion that in this case the temporal evolution of most of the stellar parameters is very similar to predictions for non-rotating stars with moderate convective core overshooting (of course the evolution of the surface CNO abundances is different, but as stated already, this effect only marginally affects the overall evolution of a massive star). Mokiem et al. (2006) investigated 21 OB type dwarfs in the SMC and came to the conclusion that the average

$v_{rot}$ = 160-190 km/s. Since these stars are dwarfs (thus close to the ZAMS) with a low metallicity (thus a small stellar wind mass loss rate), the $v_{rot}$ values should be representative for the initial value, and interestingly, this value corresponds to the value for the galactic O-type stars (Figure 1). Note that

- the similarity between the ZAMS values of the Galaxy and the SMC seems to indicate that there is no difference between the initial rotational velocity of O-type stars in the Galaxy and the SMC,
- lower initial rotational velocities for O-type stars on the ZAMS imply lower rotational velocities when the star becomes a B-type supergiant. Markova and Puls (2008) re-analyzed a number of bright supergiants in the Galaxy. They determine the rotation velocity of these stars by accounting for macro-turbulence (which has a similar line-broadening effect as rotation) and they derive values which are up the 40% lower than previous studies where macro-turbulence is not included, a result that indeed corresponds to the lower initial O-type rotation velocities defended in the present paper.

Of course, the distribution depicted in Figure 1 shows that there is a significant group of rapidly rotating O-type stars. However, a close look at this sample reveals that many of these stars are runaways with a peculiar space velocity > 30 km/s. This means that many of them do not have a canonical single star history. It is more likely that they were former binary components and became runaway due to the supernova explosion of the companion, or they became runaway due to dynamical interactions in a dense stellar aggregate. As will be discussed more in detail in the second part of this paper, such runaways are in many cases the merger of two or three stars and therefore also here the canonical single star scenario does not apply.

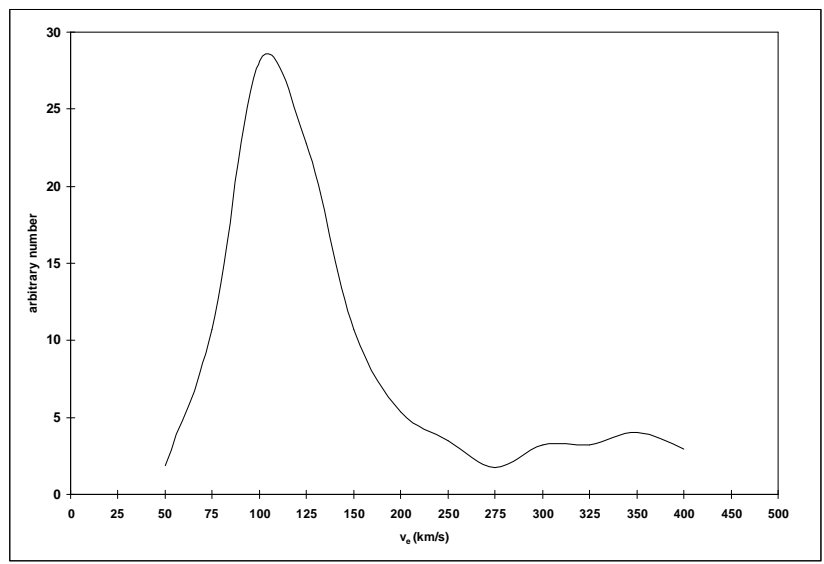

**Figure 1:** The probable $v_{rot}$ distribution of O-type stars using the data of Penny (1996) (from Vanbeveren et al., 1998)

What about the rotational properties at birth of the early B-type dwarf and giant massive star sample? Could it be that these properties are different from those of the more massive O-type sample? For the Galactic clusters NGC3293 and NGC4755 Dufton et al. (2006) obtain a mean equatorial rotational velocity for B-type stars (spectral type earlier the B9) between 225-250 km/s. High rotation velocities of B-type stars are also observed in the double cluster h and $\chi$ Persei (Strom et al., 2005). And we may obviously not forget the large population of Be-type stars rotating with a velocity close to break-up., although we know that many early Be-type stars have a neutron star companion (Be-X-ray binaries). This indicates that these Be-type stars were the mass gainers of the progenitor binaries and probably acquired their rapid rotation via mass transfer (see also section 2.4).

There are two important studies that illustrate that the evolutionary models of rotating massive stars still lack some basic physics.

- The meridional circulation related to rotation produces massive MS stars with altered surface CNO abundances. As part of the VLT-Flames survey, Hunter et al. (2008) observed about 700 massive OB stars and investigated the rotation velocities and the surface abundances of these stars. Curiously, a number of slow rotators show significant N-enhancement in the surface whereas a number of rapid rotators show no or only moderate N-enhancement, and this is not what is predicted by the rotating models of single stars. It looks as if rotational mixing in rotating single stars may be much less efficient than commonly thought. As will be discussed later on, N-enrichment in slow rotators may require a binary solution.
- Last but not least, Heger et al. (2004) computed the spin periods of neutron stars predicted by rotating massive star models. When they compared their results with observations of pulsars they concluded that the models predict neutron stars that are spinning too fast by at least an order of magnitude. Interestingly, Suijs et al. (2008) demonstrated that the evolutionary models of rotating lower mass stars produce white dwarfs that are rotating too fast as well.

**2.3. Rotationally induced magnetic fields and massive single star evolution**

As shown by Spruit (2002) a dynamo can operate in the radiative zone of a differentially rotating star. The resulting magnetic field may exert an efficient torque able to reduce the differential rotation and force the star to rotate uniformly. The effects of this process on massive star evolution have been studied by Heger et al. (2005) who concluded that (not unexpectedly) stars slow down much more rapidly and this of course reduces the overall effect of rotation on massive star evolution whereas the distribution of pulsar spin periods is much better reproduced when magnetic fields are included. Maeder and Meynet (2005) calculated the evolution of a 10 $M_\odot$ star with and without rotationally induced magnetic fields, and also their results seem to indicate that the overall influence of rotation on stellar evolution becomes smaller when magnetic fields are included.

Therefore, sticking out my neck, by considering all arguments, I am inclined at present to conclude that for cluster population and spectral synthesis simulations the effect of rotation on massive O-type single star evolution may be important but it is not very important.

**2.4. Massive binary evolution**

Various research groups all over the world have studied the evolution of massive binaries for more than 40 years. An extended review was published in 1998 (Vanbeveren et al., 1998) and in the same year our monograph 'The Brightest Binaries' summarized the state of massive binary research till then. Some subsequent studies are discussed below.

**a. The primary = mass loser**

When a massive star is the primary of a binary it may lose all its hydrogen rich layers by RLOF, and becomes a hydrogen-depleted helium-burning star, resembling a WR star. The process happens in most of the massive binaries with a period less than 10 years when the initial mass of the primary is $< 40$ $M_\odot$ (those with larger masses may encounter a LBV phase and the binary evolves through the 'LBV'-scenario as it was introduced in Vanbeveren, 1991). This means that also a 10 $M_\odot$ primary will lose all its hydrogen rich layers and becomes a WR-like star.

Due to the RLOF most of the massive primaries in interacting binaries do not become red supergiants (or have a very short red supergiant phase). This means that a population synthesis code that does not account for binaries may predict a red supergiant population that is completely wrong if in reality the population has a significant number of binaries. I think that it is important to realize that the most commonly used spectral synthesis code 'Starburst99' does not account for binaries at all.

The tidal spin-orbit process in short period binaries is very efficient and the synchronization of the rotation and orbital revolution of the components is very rapid. The effect of rotation on the evolution of massive primaries in short period binaries is therefore rather small.

**b. The secondary = mass gainer**

The evolution of the secondary in an interacting binary is affected by the RLOF of the primary. Summarizing:

- Part of the mass lost by the loser may be accreted by the gainer; the gainer becomes bluer and brighter (the rejuvenation process of the gainer)
- Accretion of mass implies accretion of angular momentum, the gainer spins up and may become a very rapid rotator (Packet, 1981).

Before 1998, the role of rotation on binary evolution was investigated in only few studies. Vanbeveren et al. (1994) and Vanbeveren & De Loore (1994) introduced the accretion induced full mixing model. The idea was the following:

- depending on the entropy and the chemical composition of the mass lost by the loser, accretion of this mass implies mixing with the gainer's interior
- accretion of mass implies accretion of angular momentum; the gainer spins-up and this implies rotational mixing.

We proposed a binary evolutionary model where the effect of rotation was simulated by assuming that after the RLOF and accretion phase, the gainer is mixed completely and then further evolves as a normal (but rejuvenated) star. This model was capable in order to explain the overluminosity and the He-overabundance of some optical components of high mass X-ray binaries.

A detailed study of the effects of rotation on the evolution of the mass gainers in massive binaries was presented by the Utrecht group (Petrovic et al., 2005; Detmers et al., 2008). Rotation is included in the stellar structure equations, angular momentum transport by internal magnetic fields is accounted for and a full spin-orbit coupling through tides decides upon the evolution of the rotation of the gainer. The following points can summarize their results.

- Rotational mixing in mass gainers can be very efficient, which allows us to conclude that our accretion induced full mixing model of 1994 was not too exotic.
- The observations of Hunter et al. (2008) which are difficult to explain by rotating single star models (see also section 2.2), can be understood in terms of binaries, mass gainers and rotation (Langer et al., 2008).
- In small metallicity regions where stellar wind mass loss is small, rapidly rotating mass gainers remain rapid rotators till core collapse and can therefore be considered as plausible models for long gamma ray bursts.

### c. Mergers

In many binaries where the RLOF is non-conservative (mass lost by the loser leaves the system) both components are expected to merge. The merging process of two stars is still poorly known but for the evolution of stars it is surely as far-reaching as any other process that is very popular nowadays. SPH simulations of massive star mergers reveal efficient mixing (Suzuki et al., 2007). Even more the merger of two binary components may be a rapid rotator increasing the efficiency of mixing. A plausible evolutionary model for mergers may be the merger induced full mixing model where both components are mixed homogeneously and 'normal' evolution starts over again.

### 3. Massive star population synthesis including a realistic fraction of binaries

The massive star population results discussed in the following relies on the Brussels code, which account for all the binary processes that are known, and include an extended data set of single and binary evolutionary models. In particular, the evolution of the mass gainers is simulated accounting for the accretion induced full mixing model (e.g., we account for the possible effects of rotational mixing due to mass accretion in mass gainers). Also the mergers are followed in detail where the merger induced full mixing model is used (see previous section). A detailed description of the code can be found in the papers listed below.

### 3.1. The rapidly-rotating-star population in starbursts

Starting with a starburst with non-rotating stars but with a reasonable fraction of primordial binaries, Van Bever and Vanbeveren (1998) calculated the temporal evolution of the population of mass gainers and mergers (= the population of rapidly-rotating-stars) in clusters younger than 11 Myr. Figure 2 shows a typical situation in the HR-diagram after 8 Myr. We conclude:

- After 4-5 Myr, blue stranglers formed by binary mass transfer and mergers dominate the luminous and hottest part of the starburst; many of them are expected to be rapid rotators. This means that a starburst which had originally only 'slow' rotators but with a fraction of primordial binaries will become populated with fast rotators because of binary evolution (mass gainers and mergers).

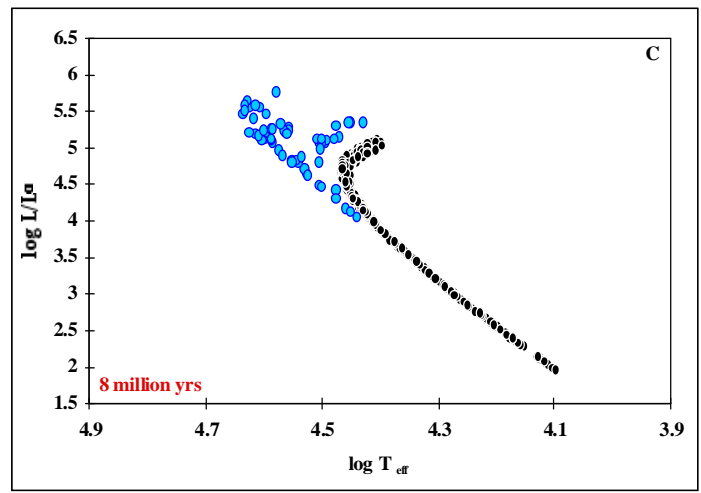

**Figure 2:** A typical starburst with primordial binaries after 8 Myr; the blue stars are rapidly rotating mass gainers or binary mergers (from Van Bever and Vanbeveren, 1998).

- The mass gainers and mergers make starbursts look younger (the rejuvenation of starbursts). This is also visible in the temporal evolution of the nebular $H_\beta$ line (Figure 3). A starburst with an equivalent width $W(H_\beta) < 100$ Å will be catalogued as younger that 5 Myr by Starburst99; with a reasonable fraction of binaries it may be as old as 10 Myr.

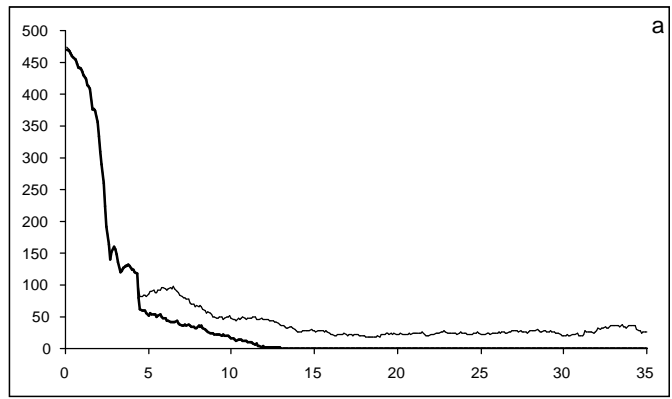

**Figure 3:** The temporal evolution of the equivalent width (in Å) of $H_\beta$ of a starburst without binaries (thick line) and with binaries (thin line). The ages on the horizontal axis are in Myr (from Van Bever and Vanbeveren, 1998)

Van Bever and Vanbeveren (1997) followed the population of mass gainers and mergers in starbursts up to 300 Myrs, mainly to compare theoretical prediction and observations of the population of Be-type stars in some Galactic and Magellanic Cloud (MC) clusters. We concluded that

- The population of rapid rotators in general, the Be-type stars in particular in the Galactic clusters h & χ Per, NGC 663, NGC3760 and in the MC clusters NGC 330, NGC 2004, NGC 1818 can not be explained by a starburst model with primordial binaries where all the cluster members were formed more or less instantaneously. The discrepancy is at least a factor 10.

### 3.2. The WR population in starbursts

The effect of binaries on the WR population and the WR spectral synthesis of starbursts has been investigated by Belkus et al. (2003). To illustrate, Figure 4 compares the temporal evolution of the WR line CIV λ1550 predicted by Starburst99 and predicted by our spectral synthesis code with a realistic fraction of massive binaries.

We conclude

- A starburst that shows WR spectral features is considered younger that 5 Myr by Starburst99, but could be twice as old if one accounts for binaries.

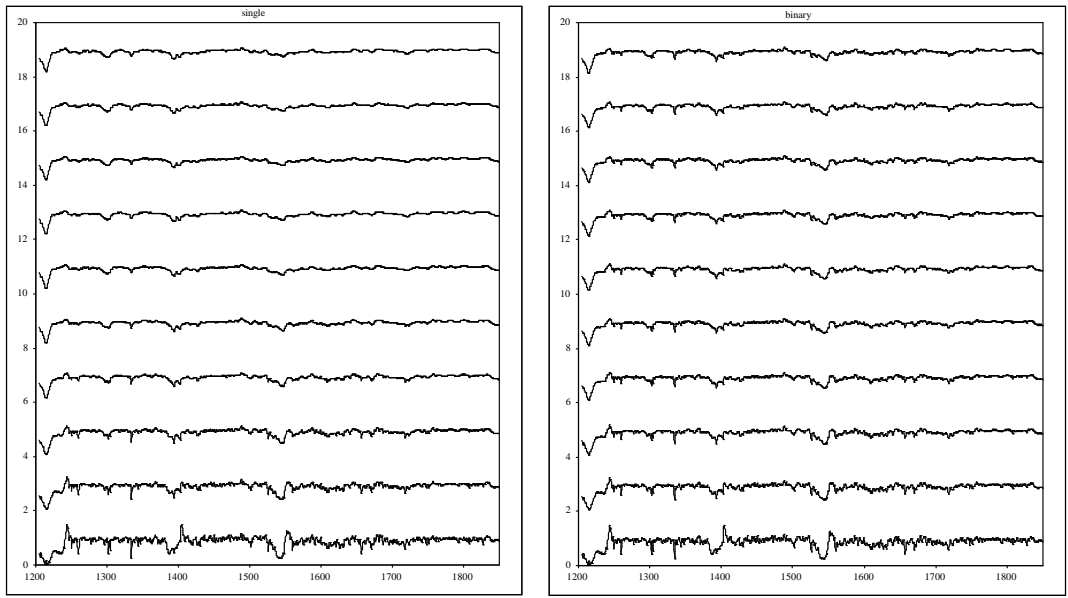

**Figure 4:** The predicted temporal evolution of the WR line CIV λ1550 in a starburst without and with binaries (from Belkus et al., 2003). The numbers on the vertical axis are Myr.

### 3.3. The X-ray binary population in starbursts

Of course only starbursts with a primordial binary population will emit after some time (typically after 3 Myr) hard X-rays produced in X-ray binaries. The X-ray binary population in starbursts has been studied in detail by Van Bever and Vanbeveren (2000).

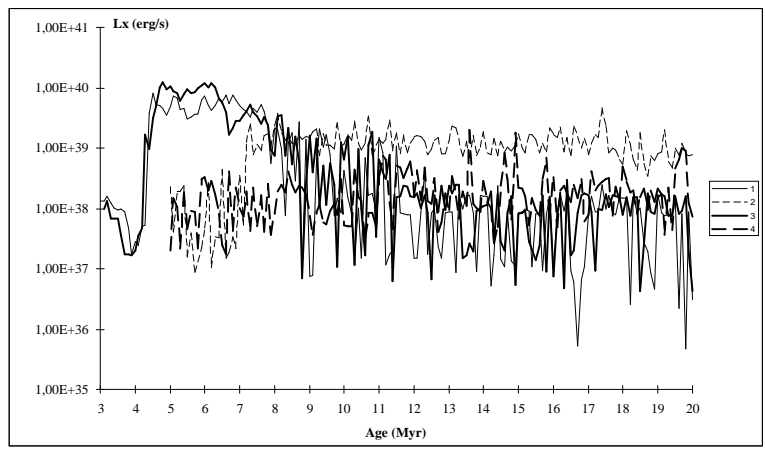

**Figure 5:** The stochastic temporal evolution of the hard X-radiation produced in X-ray binaries and young supernova remnants in a typical starburst containing a realistic population of primordial massive binaries (from Van Bever and Vanbeveren, 2000).

We follow the X-ray binaries with a neutron star companion, a black hole companion but also the X-rays emitted by young supernova remnants. A typical prediction is shown in Figure 5 but we refer to the paper for more details.

## 4. Dynamics in dense clusters: the formation and evolution of very massive stars

Using a direct N-body code (Starlab), Portegies Zwart et al. (2004) simulated the dynamical evolution of a cluster with the properties of MGG-11 in M82. The authors concluded that a runaway collision process starts in the core after less than 1 Myr and an object is formed with a mass of the order of 1000 $M_\odot$. They proposed that this object could have produced an intermediate mass black hole (IMBH) of comparable mass, leading to a sub-Eddington explanation for the ultra luminous X-ray source (ULX) observed in MGG-11.

Portegies Zwart et al. did not really consider the evolution of such a very massive star in detail, but Belkus et al. (2007) did and it was concluded that

- the answer on the question whether or not an IMBH can be formed is blowing in the stellar wind mass loss rate of very massive stars.

Figure 6 illustrates the mass evolution of very massive stars with a plausible stellar wind mass loss rate scenario. As can be seen, the very massive stars (after formation) lose most of their mass and an IMBH is not formed. Note that Yungelson et al. (2008) essentially obtained results which are similar to ours.

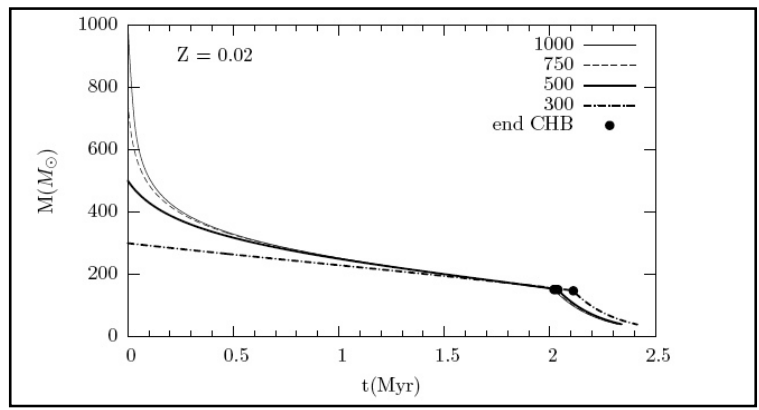

**Figure 6:** The mass evolution during core-hydrogen-burning and during core-helium-burning of very massive stars (Solar metallicity) (from Belkus et al., 2007).

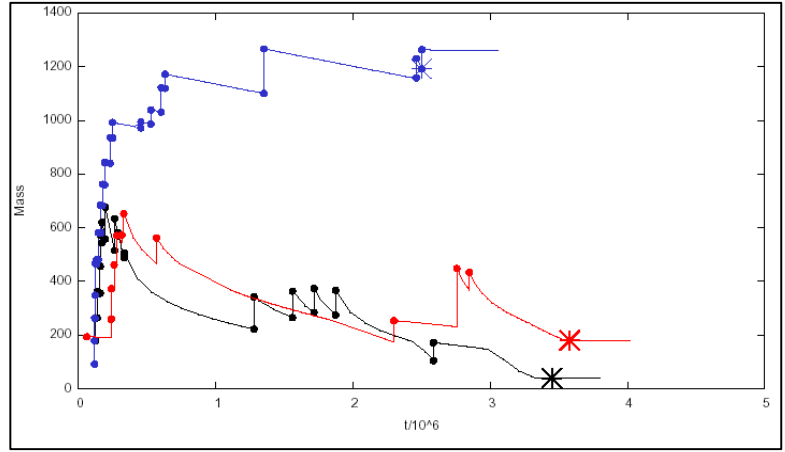

**Figure 7:** The mass evolution of the collision runaway object in a MGG-11 type cluster. Blue = simulation with small mass loss, similar to the results of Portegies Zwart et al. (2002), black = simulation with the stellar wind mass loss formalism as discussed in Belkus et al. (2007) for Solar type metallicity, red = same as black but for a SMC type metallicity (from Vanbeveren et al., 2008).

The next step was to include the evolution of very massive stars into an N-body code and re-calculate the dynamical evolution of MGG-11. Results were published in Vanbeveren et al. (2008) and typical simulations are shown in Figure 7. We conclude:

- stellar wind mass loss does not prevent the initial collision rate and a very massive object is formed. However, after this runaway event, the remaining evolutionary timescale is of the order of 2 Myr and stellar wind mass loss is capable to remove most of the mass of the star.
- The formation of an IMBH with a mass of the order of 1000 $M_\odot$ in an MGG-11 type cluster is rather unlikely.
- In an MGG-11 type cluster but with a very low metallicity and thus a much smaller efficiency of stellar wind mass loss, an IMBH may be formed with a mass of a few 100 $M_\odot$ (the red curve in Figure 7).

Leonard and Duncan (1990) presented quantitative calculations of the percentage of runaway stars (stars with a peculiar space velocity > 30 km/s) formed via close encounters in the dynamical environment of dense clusters. Also our simulations reveal the formation of such dynamically formed runaways. Interestingly, most of them are a merger of two or even three stars and are expected to be rapid rotators.

Therefore, as a final conclusion, *if on asks whether or not rotation is important for stellar evolution, the answer is yes but perhaps mainly in the framework of binary evolution (rapidly rotating mass gainers and binary mergers) or in the framework of dynamics in dense clusters where stars collide, merge and become rapid rotators*.